\documentclass[11pt]{article}
\usepackage{graphicx} 
\usepackage[
backend=biber,
style=nature,
]{biblatex}
\usepackage[parfill]{parskip}
\usepackage{authblk} 
\usepackage{amsmath}
\usepackage[left=1.5in,
  right=1.5in,
  top=1in,
  bottom=1in]{geometry} 

\usepackage{color}
\usepackage{url}
\definecolor{mypurple}{rgb}{0.82,0.02,0.48}
\definecolor{mygreen}{rgb}{0.1,0.6,0.2}
\definecolor{myblue}{rgb}{0.53,0.81,0.92}

\title{How much regulation do we need from genomes to society?} 
\author[1,2]{Vicky Chuqiao Yang}
\author[3,6]{Christopher P. Kempes}
\author[3]{S. Redner}
\author[3]{Jos\'e Ignacio Arroyo}
\author[3]{Geoffrey B. West}
\author[3,4,5]{Hyejin Youn}

\affil[1]{MIT Sloan School of Management, Massachusetts Institute of Technology, Cambridge, MA}
\affil[2]{MIT Institute for Data, Systems, and Society, Massachusetts Institute of Technology, Cambridge, MA}
\affil[3]{Santa Fe Institute, Santa Fe, NM}
\affil[4]{Seoul National University, Seoul, Korea}
\affil[5]{Northwestern Institute on Complex Systems, Evanston, IL}
\affil[6]{Complexity Science Hub, Vienna, Austria}

\date{}

\begin{document}
\maketitle
 
\section*{Abstract}
Regulatory functions are essential in both socioeconomic and biological systems, from corporate managers to regulatory genes. Regulatory functions come with substantial costs and benefits, and the balance of the two is often taken for granted. A fundamental question for all complex systems becomes how much regulatory function do they need for their size and function? Here, we present empirical evidence that regulatory functions scale systematically across diverse systems: biological organisms (bacterial and eukaryotic genomes), human organizations (companies, federal agencies, universities), and decentralized entities (Wikipedia, cities). We combine an analysis of large data sets from each of these domains with a simple conceptual model. The model predicts that the scaling of regulatory costs shifts with system structure. Well-mixed small systems exhibit superlinear scaling between size and regulatory function, while modular large ones show sublinear or linear scaling, both in agreement with data. Finally, we find that socioeconomic systems that contain more diverse occupational functions tend to have more regulatory costs than expected from the scaling relationships, confirming the hypothesis that the type and complexity of interactions also play a role in regulatory costs. Our cross-system comparison offers a mechanistic framework for understanding regulatory function and can potentially guide efforts to analyze the costs and benefits of regulatory function in diverse systems.

\section {Introduction}
Regulatory functions and mechanisms are fundamental for maintaining stability, managing complexity, and coordinating interactions in both biological and social complex adaptive systems. These processes operate across different scales, from individual cells to entire societies, and rely on dedicated components that regulate internal processes and mediate interactions among constituent parts. In biological systems, regulatory functions prevent harmful interactions between components that may be individually beneficial but detrimental when expressed together. For this, regulatory genes control the timing and conditions of gene expression, ensuring that conflicting interactions are avoided. For example, allosteric regulation of proteins, such as serpins, helps inhibit diseases such as emphysema \cite{sanrattana2019serpins}. Likewise, in human organizations, regulation is presumed to optimize efficiency and prevent dysfunction by controlling workflows, overseeing tasks, and resolving conflicts \cite{mitzberg1973, mitzberg1979}. At the societal level, legal and institutional regulations mediate disputes, allocate resources, and prevent systemic failures and thus maintain order and balance among competing interests \cite{weber1978max}.

These necessary and ubiquitous regulatory functions consume a significant amount of energy and resources. For example, regulatory genes in bacteria account for about  $10$\% of the metabolic costs \cite{lynch2015bioenergetic}. In the US, 15\% of workforce compensation is paid to managers \cite{blsWages}. In US universities, administrative spending is on par with instructional spending, and has been cited as a key factor in the skyrocketing tuition cost \cite{mumper2005causes, greene2010administrative}. In fact, the burden of administrative costs is a significant concern in many aspects of society, including higher education \cite{ginsberg2011fall}, health care \cite{himmelstein2020health}, manufacturing \cite{meyer2013limits}, and the transition to renewable technologies \cite{klemun2023mechanisms}.  As a result, regulatory costs have emerged as one of the major societal challenges of the 21st century. What aspects of regulatory function are set by fundamental requirements and which are malleable through changes in structure, culture, or procedure? The answers to these big questions require more understanding of the mechanisms underlying regulatory functions across a wide range of systems, but a unified understanding across these domains  has yet to be developed.


Historically, regulatory costs have been studied independently in the domains of biological organisms, organizations, and societies, yet a comprehensive understanding that bridges these domains has remained a longstanding open challenge. In the study of biological organisms, we have a clearer understanding of the fundamental and baseline requirements for regulatory function. A key assumption in the biological context is that the overall evolutionary process, involving vast numbers of species and multiple timescales, tends towards the optimization of regulatory costs \cite{kempes2019scales}. Thus, biological examples provide a useful case study for understanding essential or optimal regulatory functions. Furthermore, similar key determinants of size, internal structure, and complexity emerge from the biological literature. For instance, major transitions in biological architecture --- such as the evolution from bacteria, which lack internal compartments and allow any gene to interact with any other gene, to eukaryotes, which have compartmentalized cells and multiple chromosomes, to multicellular organizations --- each introduces new and distinct forms of regulation \cite{szathmary2015toward}. 

Similarly, regulatory costs in socioeconomic systems are highly complex and take many different forms, each representing an area of study, including managerial overhead, administrative intensity, and bureaucratic burdens. For example, in corporate environments, the role of managers has been extensively studied, with particular attention to best practices for improving efficiency and coordination \cite{VanReenen2010JEP}. Middle and upper management oversee workflows, resolve conflicts, and align operations with institutional goals, yet these functions also contribute to administrative overhead \cite{VanReenen2010JEP, Hjort2022, Chandler1977}. Similarly, regulatory structures in government and other institutions mediate interactions, allocate resources, and prevent systemic failures, ensuring stability within complex social systems \cite{jennings2005weber, hannan1984structural, Ostrom}.

Across these diverse contexts, regulatory mechanisms in both biological and social systems share a fundamental purpose: maintaining order and preventing or resolving conflicts \cite{malone1994coordination, herbert1947}. In biological systems, regulation operates through genetic and metabolic mechanisms that maintain cellular homeostasis and enable adaptive responses to environmental changes \cite{konieczny2023regulation}. In human organizations, regulatory functions manifest through governance, law enforcement, and institutional coordination, ensuring social stability and managing disputes \cite{march1993organizations, mintzberg1980structure, herbert1947, Ostrom}. In both cases, regulation serves to monitor activities, enforce rules, and maintain overall system stability.

There are several key determinants of regulatory costs across these domains. These include size, often measured by the number of individuals in the system, such as employees or genomes \cite{boyne2013burdened, van2003scaling}; internal structure, measured by the level of hierarchies and size of the sub-units \cite{carillo1991organization}; and functional complexity \cite{anderson1961organizational}, often measured by the number of different tasks performed by individuals. While these factors are well-documented within their respective fields, a comprehensive and systematic understanding of the determinants of regulatory costs remains elusive.



The identification of common key determinants across both social and biological systems, such as size, structure, and functional complexity, suggests the potential for a unifying and comparative perspective. A powerful tool that has been successful in accomplishing this across a diverse range of systems, including those in physics, biology, and ecology, as well as for firms and cities \cite{west2018scale,van2003scaling, kempes2017drivers,cody1975ecology,bettencourt2007growth} is the use of scaling analyses to reveal common underlying mechanisms. Fundamental metrics in many of these systems exhibit simple power law scaling behavior as a function of their size, which can be mechanistically, and in many cases quantitatively, understood from similar underlying dynamics. Such studies reveal key connections between size, function, and architecture, illustrating how fundamental principles can be applied across different types of systems. A major contribution of this paper is to show that metrics reflecting regulatory functions from bacteria to cities and companies do indeed scale in a similar and systematic fashion, strongly suggesting that there are unifying principles and dynamics at play. 
 
First, we conceptualize what gives rise to regulatory costs across complex systems based on the costs to manage adverse interactions. We compare, contrast and unify the degree to which the constituents of these diverse systems are well-mixed or highly modular. At one end of the spectrum,  well-mixed systems tend to be the simplest self-organized and agglomerated ones. At the other end, those with modular structures tend to be centrally planned or have gone through several transitions in architecture. For example, bacteria are defined by a cellular environment where most expressed proteins in the liquid cytoplasm can diffuse and interact with any other expressed protein, posing unique regulatory challenges for governing protein co-expression. At the genome level, in bacteria the genome is physically distributed in a nucleoid with no surrounding compartment, in contrast to the genome of eukaryotes which is located in the nucleus with a complex compartmentalization, that includes not just distinct chromosomes, but also territories and other structures \cite{meldi2011compartmentalization,cremer2010chromosome}. In contrast to bacteria, companies are typically defined by hierarchical and modular structures that regulate interactions among individuals and are highly planned as they grow in size. This pattern also applies to more complex organisms that have experienced multiple major evolutionary transitions in structure \cite{szathmary2015toward}. We then compile data on regulatory costs in biological and social systems, including regulatory genes in both bacteria and eukaryote cells, managers in companies, governmental agencies and universities, administrators on Wikipedia, and lawyers in 
cities. We show that all of these scale with system size as power laws, indicating that, despite the broad diversity, they are all manifestations of self-similar structures.  Their apparent differences can be quantified by the value of the scaling exponents in how their regulatory costs change with size to reveal shared governing processes and principles across different systems.



\section {Results}

We define regulatory functions as entities whose primary role is to moderate, adjust, or coordinate the interactions among other entities. These include regulatory genes in cells, managers in companies, lawyers in cities, and administrators on Wikipedia. Examples of functions that do not fall in this category include primarily functional components, for example, functional genes in cells, factory workers in a manufacturing plant, and primarily maintenance and repair functions, such as janitors of a university. 

\subsection{Model for baseline expectations of regulatory costs across systems}

\begin{figure}[htb]
    \centering
    \includegraphics[width=0.7\linewidth]{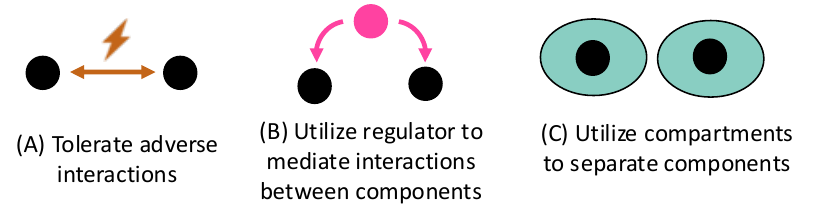}
    \caption{Conceptual illustration depicting three strategies complex systems could employ to manage adverse interactions among components. Each strategy incurs a cost, and the optimization of the total costs forms the basis of our model.}
    \label{fig:modelIllustration}
\end{figure}

We present a simple mathematical framework to illustrate how baseline requirements for regulation can be derived from the interactions of a few key mechanisms. The central premise is that regulation arises as a response to potential adverse interactions among system components \cite{malone1994coordination}. A complex system brings together a large number of individual components — such as genes in cells or employees in companies — to achieve certain benefits, like metabolic energy for cells or revenue for companies. We denote the amount of these benefits as $B$. However, adverse interactions among these components can diminish these benefits. In cells, such interactions may occur when expressed proteins interact in ways that are futile or detrimental to cellular metabolism. In organizations, adverse interactions can manifest as duplicated efforts or interpersonal conflicts between individuals. When two components of a system have an adverse interaction, we identify three responses the system can employ, as illustrated in Fig.~\ref{fig:modelIllustration}. First, the system can do nothing and tolerate the adverse interaction. Second, it can use a regulator to manage these interactions. In cells, it takes the form of carrying a regulatory gene, which makes sure the genes encoding two negatively interacting proteins are not expressed at the same time. In organizations, it can take the form of assigning a manager to coordinate tasks and prevent duplicated work between individuals or mediate interpersonal conflict between two employees. Third, the system can separate the two components by creating compartments. In cells, this involves developing internal architecture, such as mitochondria, to ensure that certain genes are only expressed in specific sub-portions of the cell. In organizations, this approach could involve structuring individuals into separate teams, units or departments, thereby modularizing their efforts. 

Each of these three strategies carries a cost. We denote costs in these three categories as: the cost of adverse interactions, $I$, the cost associated with regulators in the system, $R$, and the cost associated with compartments $C$. Regulators and compartments each reduce adverse interactions, but come with their own costs such as the pay required to employ a manager or the energy dedicated to maintaining and expressing a regulatory gene. Compartments differ from regulators in that they use structural separation, such as organizational divisions, sub-cellular membranes, or genomic regions, to isolate unnecessary interactions. It's important to note that the cost associated with regulators and compartments may interact. For instance, in companies, the establishment of a new compartment, like a new division, is typically accompanied by the appointment of a regulator, such as the division head.

\begin{figure}[htb]
    \centering
    \includegraphics[width=0.7\linewidth]{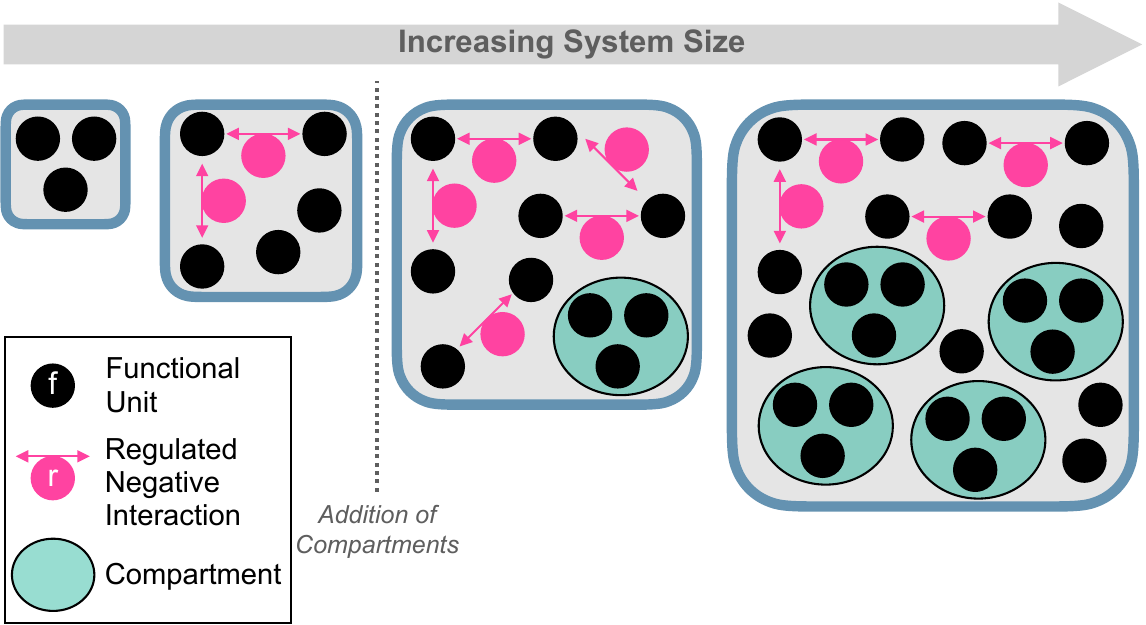}
    \caption{The scale-based model of function, regulation, and compartments. Functional elements are able to freely interact without compartments or regulators. Regulators modulate pairwise interactions and compartments restrict interactions to functional elements within the same compartment. The size of the system, $N$, is the total number of functional elements. Equation \ref{general-cost} quantifies the costs associated with negative interactions, regulators, and compartments.}
    \label{fig:modelscaling}
\end{figure}

We are interested in how these benefits and costs change with the system size, $N$, such as the number of employees in an organization or the number of genes in a genome (Fig.~\ref{fig:modelscaling}).  In particular, we specify the number of functional individuals, $f$, regulators, $r$, and the number of compartments, $c$, to arrive at the generic utility function
\begin{equation}
\label{general-cost}
\mathcal{L}= B(f,r,c)-I(f,r,c)-R(f,r, c)-C(f,r, c).
\end{equation}
The explicit expression of each term will depend on the system of interest. For example, in cells, the cost of compartments is related to the physical maintenance of these structures and, thus, is influenced by their surface area.  In organizations, these costs may be linked to the implementation of codified processes and the coordination required between departments.

Equation ~\ref{general-cost} enables the optimization of the utility function, given the analytical forms for the four terms, $B, I, R$ and $C$, as functions of the number of regulators, $r$, and the number of compartments, $c$. While the precise quantitative forms of these four terms still remain uncertain for specific systems, it is useful to introduce the simplest version of each term to understand the baseline optimization of $\mathcal{L}$ and how $r$ and $c$ scale with system size. We present a summary of our findings below, while the full derivation can be found in the Supplementary Information. 

In a well-mixed environment in which all constituent components interact equally with all others, the addition of a new functional individual---whether an employee or gene---comes with a probability $\mu$ of having a negative interaction with all existing functional members of the system. If we consider the $f$ functional individuals as identical, then the total number of negative pairwise interactions is $\mu f\left(f-1\right)/2$. Each of these negative interactions comes with a cost $\gamma_1$, and we can remove negative interactions either by adding regulators or placing individuals in compartments. We assume that individuals contained within compartments are removed from pairwise negative interactions with the rest of the system. If we place $\eta$ individuals into a compartment, and the system contains $c$ identical compartments, then the total number of negative interactions becomes $\rho \left(f - \eta c\right)\left(f - \eta c -1\right)$, where $\rho\equiv\mu/2$. For a fixed compartment size, negative interactions within a compartment simply become a fixed cost, which we will combine with compartment costs later. Additionally, if each regulator can reduce $\theta$ negative interactions, then the total number of negative interactions becomes $\rho \left(f - \eta c\right)\left(f - \eta c -1\right)-\theta r$ leading to the final utility function, 
\begin{equation}\label{eq:L}
\mathcal{L}= b f -\gamma_{1} \bigg[\rho \left(f - \eta c\right)\left(f - \eta c -1\right)-\theta r\bigg] - \gamma_{2} r - \gamma_{3} c.
\end{equation}
where $\gamma_{2}$ is the unit cost of a regulator and $\gamma_{3}$ is the unit cost of a compartment (including the cost of negative interactions and regulators within a compartment). Here we assume that there is a proportional cost of $r$ and $c$ and a linear increase in benefit associated with $f$ such that $R=\gamma_{2} r$, $C=\gamma_{3} c$, and $B=b f$, where $b$  is the productive output of an average individual. All terms are measured in dollars or energy, depending on the system of interest. It is also useful to note that the total size of the system, $N$, which is often what is measured, is the sum of functional and regulatory components, $N=f+r$. 

\paragraph{Managing Costs with Only Regulators} Later we will show that compartments are not beneficial below a certain size, so we begin by considering an optimization for a system that only has regulators. As shown in the Supplementary Information, optimizing the utility function $\mathcal{L}$ yields the optimal number of regulators in this scenario:
\begin{equation} \label{eq:r}
    r_{\rm opt} \sim \frac{f^2 \rho }{\theta}\;,
\end{equation}
illustrating that $r$ grows like $f^2$ adjusted by the ratio of the probability of negative interactions, $\rho$, to the number of interactions that each regulator can mitigate, $\theta$. Equation~\ref{eq:r} allows us to express the total size of the system in terms of $r$ alone: $N=f+r=\sqrt{\theta/\rho}\;r^{1/2} + r $. Note that this leads to two distinct regimes characterized by how $r$ relates to $N$: for small $f$ (which corresponds to small $r$, given Eq.~\ref{eq:r}), $r \propto N^2$,  while for large $f$ (and correspondingly for large $r$), $r \propto N$. 

The first case describes an ideal scenario, most closely approximated by bacteria. There are two ways in which segregation happens in cells, either through the physical separation of expressed proteins, or through the genomic groupings of genes that are co-expressed and/or co-regulated. Bacteria have genes that generally interact in an all-to-all environment of the cytoplasm, and they do not have genes that are segregated in different chromosomes. Thus, we expect bacteria to be closest to the quadratic scaling of systems without compartments. The literature report superlinear exponents as high as $1.86$ (see Supplementary Information for a review). However, bacteria are known to shift scaling toward the largest and smallest cells for a variety of biophysical and physiological reasons that lead to asymptotic behavior \cite{kempes2012growth,kempes2016evolutionary,gondhalekar2023scaling, supo2025percolation}. It is also the case that large bacteria acquire effective compartments that occur in several ways. First, proteins are often transported to a specific location, such as the membrane to perform functions and this effectively isolates them from other proteins. Second, recently it has been shown that liquid-liquid phase separation produces a variety of membrane-less organelles in bacteria \cite{jin2021membraneless}. Third, despite having a single chromosome, bacterial genomes are partially compartmentalized using operons (clusters of co-expressed genes), and higher-order modulons (groups of operons) \cite{osbourn2009operons}. Thus, we should expect larger bacteria to deviate from quadratic scaling due to the addition of compartments. We perform a breakpoint analysis on the binned data \cite{gondhalekar2023scaling} and find two distinct scaling relationships, where the first gives the exponent of $1.65$for small cells, and the second gives $1.37$ for the largest bacteria. It is important to note that the first scaling exponent excludes the smallest cells where only a small amount of data is available and where they demonstrate their own scaling regime (see Supplementary Information). The deviation of $1.65$ from quadratic is likely due to the genomic co-location of sets of functions. The deviation of $1.37$ from linear is likely due to the more imperfect and more informal compartments of large bacteria.

For the second case, where $N\propto r$, there is a critical size at which regulators overwhelm the system causing $\mathcal{L}=0$, which occurs at $f_{r}=1+b \theta/(\gamma_{2} \rho )$. Here, $f_r$ is the maximum number of functional elements for a system with regulators but no compartments. This upper bound depends on the ratio of the benefit per functional element to an effective cost, namely, the unit cost of a regulator times the likelihood of a negative interaction. The ratio of $b \theta$ to $\gamma_{2} \rho$ could be quite large given that the probability of negative interactions could be small and the cost of an average individual, including regulators, $\gamma_{2}$, should be small relative to the productive output, $b$, of an average individual. In addition, a regulator may be able to handle many interactions so that $\theta$ could be much larger than 1. For example, if $\rho=0.10$, $\gamma_{2}=b/2$, and $\theta=10$, then $f\approx 200$, implying that an organization that handles negative interactions with only regulators could reach a reasonable size without introducing divisions or compartments. 

\paragraph{Including Compartments} As discussed above, $f_{r}$ is the maximum number of functional elements that can operate optimally in a system with only regulators; beyond this number, compartments are needed. In Supplementary Information, we optimize the utility function $\mathcal{L}$ with respect to both number of compartments $c$, and number of regulators $r$, and find that the optimal number of compartments, $c_{\rm opt}$, is
\begin{equation}
      c_{\rm opt} \sim \frac{f}{\eta}
\end{equation}
This scaling comes with a variety of cost and size requirements that we discuss in detail in the Supplementary Information. 
These results show that in addition to $c \sim f$ the optimal regulators follow $r\sim f$. Because of this linear scaling, the total number $N=f+r$ is also linear: $r\propto N$ and $c \propto N$. In this regime, the quadratic requirements of negative pairwise interactions are handled by a combination of compartments, and regulators in those compartments, each of which scales linearly. However, this solution only occurs after the transition where compartments become inexpensive enough to be viable. 

This prediction of linear scaling is supported by the observation that in unicellular eukaryotes, the number of regulatory genes, is indistinguishable from linear \cite{van2003scaling}. Unlike bacteria, unicellular eukaryotes have various internal spatial partitions, and the partitioning of genes between the nucleus and mitochondria along with the separation of genes into different chromosomes and with a complex system of regulation that is still being uncovered \cite{fritz2019chromosome,li2018gene,cremer2010chromosome}. The transition from prokaryotes to eukaryotes illustrates how the internal structure of organisms, hierarchy, and partitioning can alter the requirements for regulation.

Considering the transitions across systems of different sizes, and that the cost per individual likely varies with system size, our model suggests that the number of regulators generalizes to
\begin{equation} \label{eq:scaling}
r \sim r_{0}N^{\beta}, 
\end{equation}
where $r_0$ is the minimum cost for the smallest size ($N=1$), and $\beta$ is the scaling exponent. This result is in keeping with the general observation that other metrics in these systems not directly related to regulation follow such power-law behavior with non-trivial exponents. Our model presented above provides base-line expectations for the values of $\beta$,  linking its value to system organization: $\beta = 2$ for a compartment-free, well-mixed system, and $\beta = 1$ for a system with both regulators and compartments. Consequently, we can interpret observed exponents that are closer to $2$ as being well-mixed and compartment-free. Exponents that are close to $1$ indicate a system optimized on compartments with a mixed strategy. An exponent significantly larger than $1$ and significantly smaller than $2$ indicates the transition from regulators only to some compartmentalization. Exponents significantly less than $1$ indicate the regime where compartments are expanding and regulators are becoming cumbersome and saturating, or where there are certain economies of scale in the unit costs.

\subsection{Empirical results}

\begin{figure}[htb]
	\centering
	\includegraphics[width= 1\textwidth]{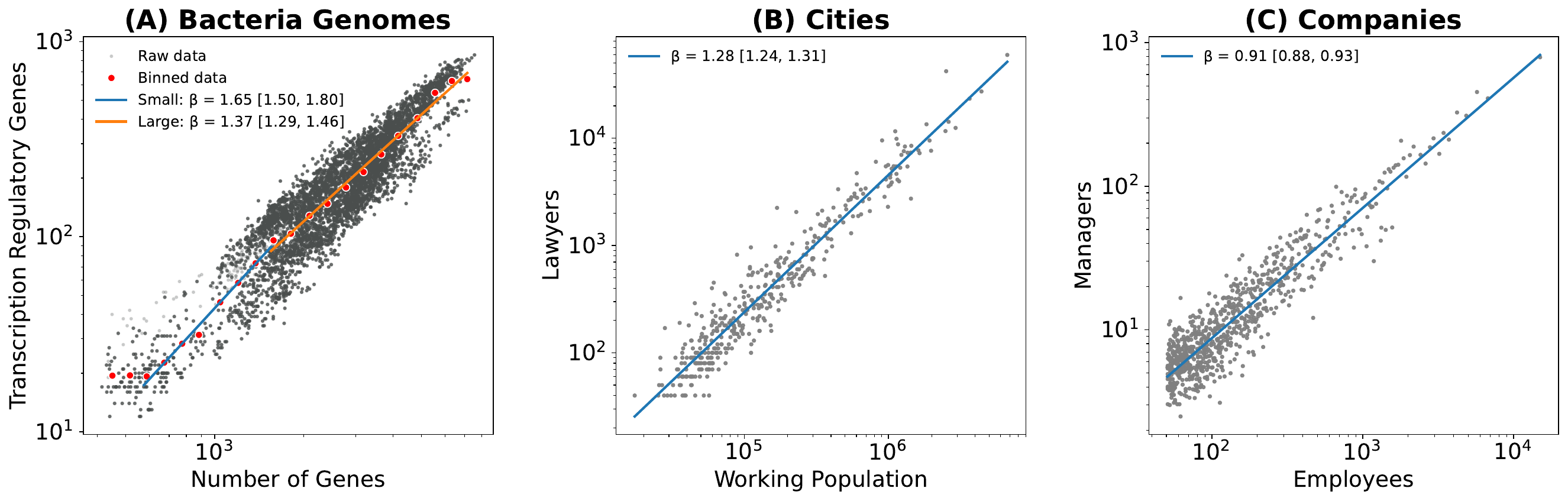}
	\caption{Examples of scaling of regulatory functions in bacterial genomes, cities, and Norwegian companies (a) The number of transcription regulatory genes vs. the number of genes in the genome of bacterial cells.  (b) The number of lawyers vs. working population in US Metropolitan Statistical Areas. (c) The number of managers in companies in Norway. }
 \label{fig:coordinationScaling}
\end{figure}
We collect data on regulatory components and system size in both biological and socioeconomic systems to examine how regulatory costs scale with size. In biological systems, we focus on bacteria and unicellular eukaryotes, specifically collecting data on the number of transcription regulatory genes in their genomes. Transcription regulatory genes are those that determine the timing and environmental conditions for gene expression by producing proteins that bind to other genes. We use data reported by \cite{van2003scaling} for these biological systems (see Supplemental Information for a comprehensive discussion). For socioeconomic systems, we gathered data on the number of lawyers in cities and the number of managers in Norwegian and Korean companies, US federal government agencies, and various types of US universities \cite{youn2016scaling, gulati2023}. We also synthesize results from another study measuring the number of administrators for Wikipedia pages \cite{yoon2023makes}. For human systems, system size is measured by the population of the entity, such as the number of employees for companies and universities, and the number of editors for Wikipedia pages. The scaling exponents of these systems are estimated using Eq.~\ref{eq:scaling},  with $\beta$ representing the scaling exponent. For detailed information on data sources and statistical methods used for estimation, see Supplementary Information. 

Based on the mathematical model, we expect the scaling of regulatory functions to vary according to the system's structure. Bacteria, representing the most well-mixed end of the spectrum, operate like a ``soup,'' where any gene can interact with any other gene. Cities, while still relatively well-mixed, exhibit a lesser degree of mixing due to the bottom-up emergence of social network clustering. We anticipate that these systems will have higher, superlinear exponents. On the modular end of the spectrum are human organizations such as companies and universities, which predominantly feature strong departmental structures. We expect these entities to have scaling exponents close to linear.

In the data we gathered, regulatory costs scale with system size across diverse systems following similar power law behavior remarkably closely. Three examples are shown in Fig.~\ref{fig:coordinationScaling}: regulatory genes in bacteria, lawyers in cities, and managers in Norwegian companies. As predicted, the scaling exponents vary across system types, with regulatory genes in small bacteria having the highest exponent at $1.65$. The number of lawyers scales superlinearly with the urban population, with an exponent of $1.28$ (Fig.~\ref{fig:coordinationScaling}B). This is similar to prokaryotic regulatory genes but with a lower scaling exponent, likely due to modularization in the interaction network structure of cities. Theories of urban scaling based on these constraints have successfully predicted similar exponents for many socioeconomic outputs driven by interactions \cite{bettencourt2013origins, yang2019modeling}. In contrast to cities, other human organizations---such as government agencies, companies, and universities---typically exhibit a high degree of hierarchical structure, leading us to expect different scaling exponents closer to unity. In these organizations, managers play a crucial role in coordinating efforts and mitigating conflicts among subordinates. Thus, the scaling exponent for managers in Norwegian companies is 0.91 in good agreement with our prediction (Fig.~\ref{fig:coordinationScaling}C).

The scaling exponents for all data we have gathered, including the number of managers in all sectors of US universities, Norwegian and Korean companies, and US federal government agencies, as well as those in biological systems, are summarized in Table~1. In all the modular systems, the number of managers scales sublinearly with small variations---from $0.94$ for Federal agencies (highest exponent) to $0.72$ for Korean companies (lowest exponent). The sublinear scaling shown in the data suggests the span of control, the number of subordinates per manager, increases with organization size, which aligns with previous studies in management science \cite{boyne2013burdened}. This finding is contrary to the popular belief that larger organizations are less efficient in coordination costs measured in managers per capita \cite{williamson1967hierarchical}; in fact, the data suggests they are more efficient at exploiting economies of scale. Even in Wikipedia, a supposedly decentralized system, the interaction network among editors has been shown to be highly modularized \cite{yoon2023makes}. Interactions tend to cluster around specific topics of interest, allowing a smaller number of administrators to effectively manage issues that arise. As such, despite the decentralized nature of the system, administrators naturally oversee specific modules and thus scale sublinearly with the total number of contributors.


\begin{table}[htb]
	\caption{The scaling exponents ($\beta$) and 95\% confidence intervals for regulatory functions in various systems. With more well-mixed internal structures, such as bacterial cells and cities, regulatory functions tend to be superlinear. With more modular internal structures, such as companies and universities, the scaling exponents are linear to sublinear.}
    \centering
	\includegraphics[width= 1\textwidth]{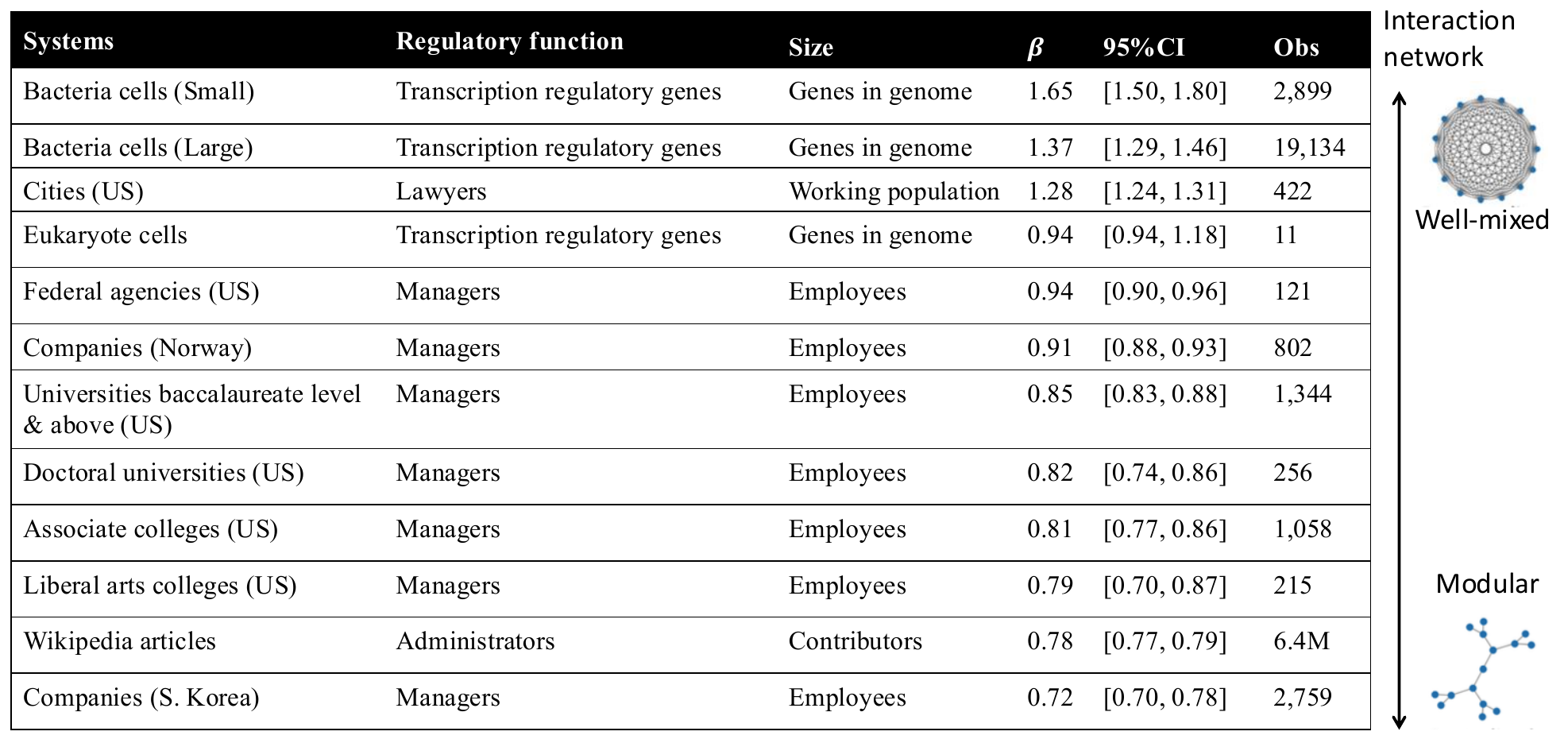}
\end{table}\label{tab:exponentTable}

\subsection{Function diversity associated with scaling deviations}
We have formulated a conceptual framework to understand how size and structure influence the regulatory costs of systems and collected data to compare with this theory. The differences across the spectrum from well-mixed to modular systems indicate regulation may arise from the number and type of interactions across individuals. We further investigate this idea by looking at how the number of regulators is related to the function diversity of a system. 

Functional diversity reflects the range of tasks performed by the components of a system. As society advances, its technology becomes more complex, and individual roles and functions become more specialized and diversified. A classic and prominent example is car assembly, which now requires specialized metal compounds for catalytic converters, computer chips, and software to manage many aspects of automobile operation. None of these components existed a half-century ago. Such an increasingly complex manufacturing operation necessitates the coordination of a broader range of components, leading to a higher potential for adverse interactions that the system needs to manage.

\begin{figure}[htb]
	\centering
	\includegraphics[width= .9\textwidth]{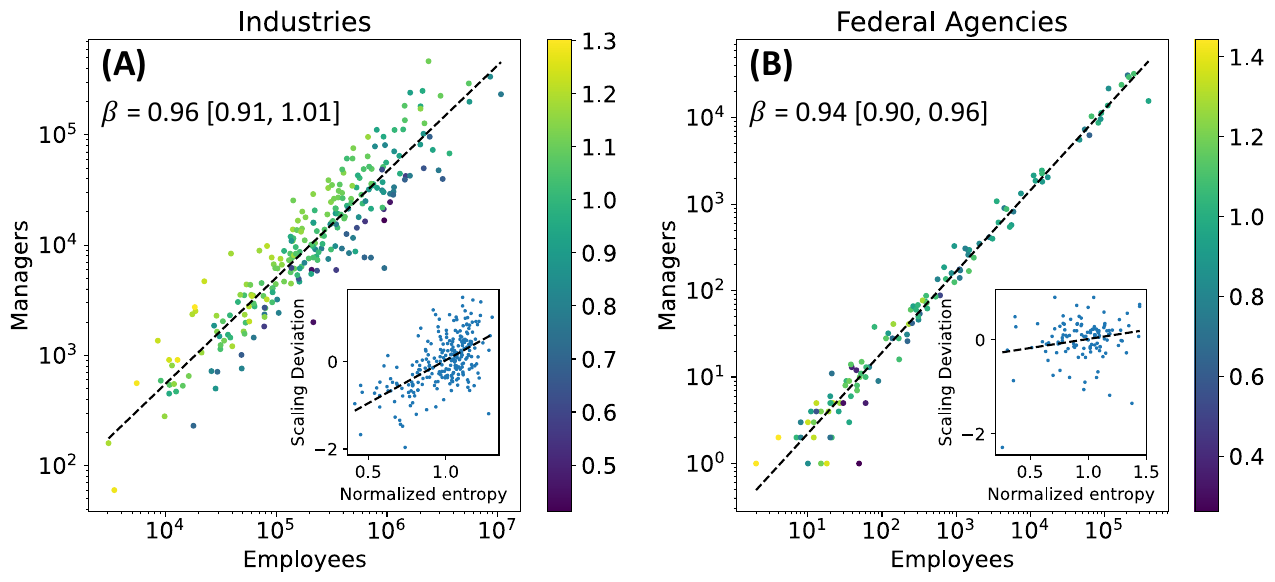}
	\caption{Data for Industries and Federal agencies colored by occupations' normalized entropy, a metric for the diversity of functions in the system. Higher values denote more diversity. The inset shows the correlation between the normalized entropy and scaling deviation. Entities that are more diverse have more managers than would be expected of their size. The Pearson correlation is 0.59 for industries and  0.20 for federal agencies.}
 \label{fig:entropy}
\end{figure}

Motivated by our theory, we predict that greater functional diversity is associated with higher regulatory costs. To test this prediction, we quantify functional diversity across a range of systems and demonstrate that it is positively associated with the scaling residuals. In other words, systems with higher functional diversity exceed the expected number of regulators as would predicted by the scaling curve, while those with lower functional diversity fall short.

We quantify functional diversity in a system by analyzing the distribution of occupations within it. For this analysis, we utilize individual-level occupation information for US federal government agencies. To ensure robustness, we also test our predictions on companies. Although occupation information is not available to us at the company level, we study companies aggregated into industries, for which we have detailed accounts of the distribution of occupations. We measure functions using the finest occupational categories available in these datasets. For more details on data sources and occupation definitions, see Supplementary Information.

We measure functional diversity using normalized Shannon entropy ($H$), an information-theoretic measure that quantifies the predictability of a function given all functions in the system. This measure has been successfully utilized to quantify diversity in socioeconomic systems \cite{frank2018small}. Mathematically, it is defined as:
\begin{equation}
    H = -\sum_{i = 1}^D p_i \frac{\log p_i}{\log D}\
\end{equation}
where $p_i$ is the relative frequency of function (occupation) $i$, $p_i = f_i/\sum{f_i}$, and $f_i$ is the frequency of function $i$. The variable $D$ is the number of distinct functions in the system. $H$ is maximized when the abundance of functions follows a uniform distribution. The normalization by $\log D$ allows comparison across systems with different total numbers of functions. 



We also compute the scaling residual \cite{bettencourt2010urban} for each system, which quantifies the extent to which a system over- or under-performs relative to the scaling curve. The scaling residual for entity $i$, $\xi_i$, is defined as, $\xi_i = \log r_i - \log r(N_i) $, where $r_i$ is the number of regulators for entity $i$ in the data, and $r(N_i)$ is the regulators expected of its size according to Eq.~\ref{eq:scaling}. 

Figure~\ref{fig:entropy} illustrates the relationship between managers and employees, with each entity colored according to its normalized entropy, where higher values indicate greater function diversity \cite{yang2022scalinguniversalityfunctiondiversity}. For both federal agencies and industries, entities with higher function diversity tend to have more managers than expected for their size. The insets of Fig.~\ref{fig:entropy} display the correlation between scaling residuals and normalized entropy. The correlation is 0.59 for industries ($p< 0.001$), and 0.20 for federal agencies ($p= 0.030$). These results support our initial hypothesis that greater function diversity is associated with greater regulatory costs as a result of greater coordination requirements. 


\section {Discussion}
We have proposed a conceptual framework for unifying the conceptualization of regulatory costs across biological and socioeconomic systems by examining the interactions of fundamental features, size and internal structure. We also conducted an empirical analysis of regulatory costs across these systems. Our findings indicate that the variation of regulatory functions with system size depends significantly on the system's structure. The exponents range from 1.65 for regulatory genes in bacteria, which have a well-mixed internal structure and genes located on a single chromosome, to sublinear scaling for managers in hierarchically organized human organizations, such as federal agencies and universities. By characterizing regulatory functions based on their underlying mechanisms rather than the system type, our work represents a first step toward a unified understanding of regulatory costs across diverse systems.

Our conceptual model is based on the simplest assumptions about cost factors, intended to lay the groundwork for more advanced theories in the future. These future models should incorporate more accurate cost estimates tailored to different system types through empirical measurements. For example, in hierarchical human organizations, the cost of regulators within compartments should also be factored into future models. In biological systems, more precise cost calculations should consider the frequency of gene expression, which plays a critical role in determining the energetic cost in the cell \cite{kempes2016evolutionary,kempes2017drivers,lynch2015bioenergetic}. 
  
It is critical to note that the greater scaling exponents of biological systems do not imply that biological systems are less efficient than socioeconomic ones. Our framework shows that for small systems, compartments are too expensive, and it is better to simply add regulators in a superlinear fashion. 
Compartmentalization and structure in biology is expensive---intricate physical structures need to be developed and maintained and this is only beneficial after a critical size. Extending this idea, complex physical infrastructure also needs to be developed to enable organs in animals, where genes only get expressed in a certain tissue. While bacteria need to carry a rapidly increasing number of regulatory genes with increasing size, they save on the energetic cost of creating compartmentalization within the cell. In comparison, developing structure in social systems does not necessarily incur a physical cost. For example, a company's CEO can decide to create a new division in the company without employing new physical separations between divisions, and the re-organization can be accomplished in a matter of weeks. Humans are also naturally creatures of groups with limited social capacity, making modularity a common characteristic in many social systems, even decentralized ones such as cities. Indeed, the social interactions scale with $N^{1.2}$ according to phone network data \cite{schlapfer2014scaling}, which is far from the $N^2$ null prediction for a completely well-mixed group.

Furthermore, superlinear exponents, such as observed in bacteria, impose a fundamental constraint on system size. With superlinear exponents, there is a maximum size beyond which all components would be regulatory. For organisms to grow beyond this limit, they must fundamentally transform their organizational structure to one that leads to a lower scaling exponent. While the superlinear exponent is inefficient for scaling up, it also has significant benefits. New genes can be easily added to the bacterial genome along with new regulation in a plug-and-play manner that provides remarkable flexibility to adapt to novel and changing environments. This insight may be transferred to small human organizations, such as start-up companies and local communities. These small organizations are similar to bacteria in the sense that new functions can be easily added with the caveat that everyone interacts directly with everyone else, leading both to a high degree of flexibility and unexpected conflicts. Our theory predicts this form of organization is limited by a predictable critical size. While this study did not collect data on start-up companies, it would be valuable for future research to analyze the regulatory costs and structure of start-up companies on a large scale and examine their transition to a more modular configuration.

Social systems, aided by the lower cost of compartmentalization, appear to gain an economy of scale with regulatory costs. However, many have the experience that larger organizations are more bureaucratic. These two observations do not necessarily contradict each other. The personal experience of regulation may reflect the experience of a non-regulatory employee complying with the structure and processes put in place by an organization. Future research should consider the cost of regulatory compliance in organizations and ask how this is traded off against regulators and compartments. It has also been noted that many regulatory costs have increased over time in many forms of organizations, such as universities \cite{ginsberg2011fall}. Temporal and cross-sectional scaling behavior differ in many socioeconomic systems, the difference can be due to changes in the output in the whole system regardless of size \cite{bettencourt2020interpretation}. Future research should extend our cross-sectional data gathering and theoretical analysis to a temporal one, to address why regulatory costs have grown in many sectors over time. 

While our study makes predictions based on structure and focuses on measuring regulatory costs across system types, we have not formally quantified the degree of modularity in the systems' architecture. Our work uses qualitative accounts of these systems. It is an area of important future work to perform a careful quantitative assessment of modularity across a wide range of system types. This would also involve gathering detailed interaction network data and quantifying the modularity of these networks.

Our work provides a unified, first-principles framework for understanding regulatory costs across biological and socioeconomic systems. In contrast to most studies in management and organizational science, which rely on regression-based analyses of correlates to regulatory burden, we develop an optimization-based model grounded in fundamental mechanisms by which systems manage adverse interactions: toleration, regulation, or compartmentalization. This model yields baseline expectations for regulatory costs as a function of system size and internal structure.

A central innovation is our cross-system comparison, linking biological and human systems through shared structural constraints. While prior work has studied regulation within each domain separately, we show for the first time that regulatory functions---from regulatory genes in genomes to managers in organizations---follow systematic scaling relationships with size, shaped by their internal architecture. This cross-system comparison can bring insights from biological systems to help understand social systems. In human organizations, regulatory costs have been rising, yet it is unclear whether such costs are an unavoidable consequence of organizational scale and complexity or the result of mismanagement and redundancy. By drawing on biological systems, which have been shaped by evolutionary optimization and where necessary regulation is more clearly defined, we uncover general tradeoffs that apply across systems, and can be used to understand essential versus excessive in human organizations.

We also show that deviations from expected scaling can be predicted by functional diversity, indicating that interaction complexity drives additional regulatory needs. Together, these contributions move beyond domain-specific explanations and toward a general, mechanistic science of regulation across complex systems.



\section*{Acknowledgements}
This research was supported by the National Science Foundation Grant Award Number EF--2133863. The authors thank Seoul Lee for helping collect and analyze South Korean company data. H.Y.~acknowledges the NRF Global Humanities and Social Sciences Convergence Research Program (2024S1A5C3A02042671) and the support from the Institute of Management Research at Seoul National University. 

\section*{Code and Data Availability}
The data used in this study and code used to analyze the data are included as a replication file with this submission. The Supplementary Information details method for data collection.

\printbibliography

\end{document}